%% file: Driver_Seismic_Chimayo.tex
\documentclass[11pt,reqno]{amsproc}
\usepackage[colorinlistoftodos,shadow]{todonotes}
\usepackage[T1]{fontenc}
\usepackage{amsmath,amscd,amssymb}
\usepackage{cite}
\usepackage{color}
\usepackage{graphicx}
\usepackage{psfrag,epsfig}
\usepackage{multirow}
\usepackage{rotating}
\usepackage[letterpaper,margin=0.75in]{geometry}
\usepackage{subfigure}
\usepackage{enumerate}
\usepackage{comment}
\usepackage{lineno}
\usepackage{algorithmic}
\usepackage{algorithm}
\usepackage[small]{caption}
\usepackage{bbold}
\usepackage[debug=false, colorlinks=true, pdfstartview=FitV, linkcolor=blue, citecolor=blue, urlcolor=blue]{hyperref}
\numberwithin{equation}{section}
\usepackage{chemarr}
\usepackage[version=3]{mhchem}
\newcommand*\patchAmsMathEnvironmentForLineno[1]{%
  \expandafter\let\csname old#1\expandafter\endcsname\csname #1\endcsname
  \expandafter\let\csname oldend#1\expandafter\endcsname\csname end#1\endcsname
  \renewenvironment{#1}%
     {\linenomath\csname old#1\endcsname}%
     {\csname oldend#1\endcsname\endlinenomath}}%
\newcommand*\patchBothAmsMathEnvironmentsForLineno[1]{%
  \patchAmsMathEnvironmentForLineno{#1}%
  \patchAmsMathEnvironmentForLineno{#1*}}%
\AtBeginDocument{%
\patchBothAmsMathEnvironmentsForLineno{equation}%
\patchBothAmsMathEnvironmentsForLineno{align}%
\patchBothAmsMathEnvironmentsForLineno{flalign}%
\patchBothAmsMathEnvironmentsForLineno{alignat}%
\patchBothAmsMathEnvironmentsForLineno{gather}%
\patchBothAmsMathEnvironmentsForLineno{multline}%
}

\title[Machine Learning for Geyser State Identification]{Using Machine Learning to Discern Eruption in Noisy Environments: A Case Study using $\mathrm{CO}_2$-driven Cold-Water Geyser in Chimay\'{o}, New Mexico}
\author[B.~Yuan et al.]{B.~Yuan$^{1}$, Y.~J.~Tan$^2$, M.~K.~Mudunuru$^{3,*}$, O.~E.~Marcillo$^{3}$, A.~A.~Delorey$^{3}$, P.~M.~Roberts$^{3}$, J.~D.~Webster$^{3}$, C.~N.~L.~Gammans$^{3}$, S.~Karra$^{3}$,  G.~D.~Guthrie$^{3}$, and P.~A.~Johnson$^{3}$ \\ \\
{\small $^{1}$Department of Mathematics, University of California Los Angeles, Los Angeles, CA 90095.} \\
{\small $^{2}$Department of Earth and Environmental Sciences, Columbia University, New York City, NY 10027.} \\
{\small $^{3}$Earth and Environmental Sciences Division, Los Alamos National Laboratory, Los Alamos, NM 87545.} \\
}
\thanks{$^*$Corresponding author, \texttt{maruti@lanl.gov}}
\date{\today}
\begin{document}
\maketitle
%
%
%
%
\section*{ABSTRACT}
We present an approach based on machine learning (ML) to distinguish eruption and precursory signals of Chimay\'{o} geyser (New Mexico, USA) under noisy environments.
This geyser can be considered as a natural analog of $\mathrm{CO}_2$ intrusion into shallow water aquifers. 
By studying this geyser, we can understand upwelling of $\mathrm{CO}_2$-rich fluids from depth, which has relevance to leak monitoring in a $\mathrm{CO}_2$ sequestration project.
ML methods such as Random Forests (RF) are known to be robust multi-class classifiers and perform well under unfavorable noisy conditions.
However, the extent of the RF method's accuracy is poorly understood for this $\mathrm{CO}_2$-driven geysering application.
The current study aims to quantify the performance of RF-classifiers to discern the geyser state.
Towards this goal, we first present the data collected from the seismometer that is installed near the Chimay\'{o} geyser.
The seismic signals collected at this site contain different types of noises  such as daily temperature variations, seasonal trends, animal movement near the geyser, and human activity.
First, we filter the signals from these noises by combining the Butterworth-Highpass filter and an Autoregressive method in a multi-level fashion.
We show that by combining these filtering techniques, in a hierarchical fashion, leads to reduction in the noise in the seismic data without removing the precursors and eruption event signals.
We then use RF on the filtered data to classify the state of geyser into three classes -- remnant noise, precursor, and eruption states.
We show that the classification accuracy using RF on the filtered data is greater than 90\%.
We also evaluate the accuracy of other classical time-series methods such as Dynamic Time Warping (DTW) on filtered data along with RF on partially-filtered data where we remove the seasonal trends.
Classification accuracy shows that DTW performs poorly (44\%) and RF on partially-filtered data performs decently (87\%).
Denoising seismic signals from both seasonal trends and human activity enhances RF classifier performance by 7\%.
These aspects make the proposed ML framework attractive for event discrimination and signal enhancement under noisy conditions, with strong potential for application to monitoring leaks in $\mathrm{CO}_2$ sequestration.
\newline
\newline
\textbf{Keywords:}~$\mathrm{CO}_2$-driven cold-water geysers,
carbon sequestration,
monitoring,
seismicity,
signal processing,
precursor,
eruption,
machine learning,
event classification,
feature extraction,
random forests.
%
\input{Sections/S1_Chimayo_Intro}
%
\input{Sections/S2_Chimayo_Method}
%
\input{Sections/S3_Chimayo_Results}
%
\input{Sections/S4_Chimayo_Conclusions}

\section*{ACKNOWLEDGMENTS}
The authors thank the support of the LANL Laboratory Directed Research and Development Directed Research Award 20170004DR.
BY and YJT like to thank the support of LANL Applied Machine Learning Summer School Fellowship. 
MKM gratefully acknowledges the support of LANL Chick-Keller Postdoctoral Fellowship through Center for Space and Earth Sciences (CSES) and UC/LANL Entrepreneurial Postdoctoral Fellowship through Richard P.~Feynman Center for Innovation.
MKM, BY, and YJT thank Youzuo Lin for many useful discussions.
MKM also thanks Ting Chen, Molly Cernicek, and Don Hickmott for their inputs and feedback during the course of the project.
Additional information regarding the seismic datasets can be obtained from the corresponding author.

\bibliographystyle{unsrt}
\bibliography{Master_References/Master_References}

\input{Sections/Chimayo_Figures}  
\end{document}

%% file: Sections/S1_Chimayo_Intro.tex

\section{INTRODUCTION}
\label{Sec:S1_Chimayo_Intro}
There is a substantial concern over the potential impact of groundwater resources in the case that a $\mathrm{CO}_2$ sequestration reservoir were to leak \cite{carroll2014key,keating2010impact}.
To a certain extent, theoretical and/or lab-scale studies can be used to predict leakage scenarios.
However, for credible risk assessments, field-scale observations of $\mathrm{CO}_2$ flowing through shallow groundwater aquifers are needed \cite{keating2013co2,keating2011challenge}.
Growing interest in collecting field-scale data for quantifying $\mathrm{CO}_2$ leakage has drawn attention to $\mathrm{CO}_2$-driven cold-water geysers \cite{keating2011challenge}.
Due to the high velocity of $\mathrm{CO}_2$-rich fluid discharge, these geysers can be thought of as a natural analog to $\mathrm{CO}_2$ leakage in a carbon sequestration project. 
$\mathrm{CO}_2$-driven cold-water geysers are similar to thermally-driven geysers (such as the ones in Yellowstone) as they have a conduit to release $\mathrm{CO}_2$-rich fluids.
While thermally-driven geysers \cite{manga2006seismic,hurwitz2017fascinating} have naturally existing conduits, in the case of $\mathrm{CO}_2$-driven cold-water geysers, the conduits are man made (typically a wellbore).
There are various field-sites of $\mathrm{CO}_2$-driven cold-water geysers across the world \cite{glennon2005operation}.
Examples of some cold-water geysers in the USA include Crystal geyser in Utah, Tenmile geyser in Utah, and Chimay\'{o} geyser in New Mexico.
The basic mechanism for $\mathrm{CO}_2$ release from these geysers is through gas expansion.
$\mathrm{CO}_2$ (gas) evolves by the pressure reduction of $\mathrm{CO}_2$-rich fluids.
Once the internal pressure of $\mathrm{CO}_2$ (aqueous) becomes greater than that of the surrounding fluid, $\mathrm{CO}_2$ (gas) separates from the fluid causing bubbles to nucleate, grow, and coalesce.
Pressure reduction occurring from increasing $\mathrm{CO}_2$ gas volume fraction enhances expansion
of $\mathrm{CO}_2$ bubbles, finally leading to an eruption \cite{watson2014eruption}.

There are various advantages of analyzing the field-data of $\mathrm{CO}_2$-driven cold-water geysers.
First, predicting when a geyser erupts is analogous to $\mathrm{CO}_2$ leak detection \cite{keating2011challenge}.
Hence, forecasting eruption events of a geyser can help in developing mitigation strategies in the case of leaks developing at a $\mathrm{CO}_2$ sequestration site.
Second, to quickly identify $\mathrm{CO}_2$ leaks in large areas, one needs to discern leakage signals from anthropocentric noise.
Due to a lot of human activity around these geysers, the sensor data serves as a perfect proxy to extract useful information from such noise.
Third, developing event discrimination capabilities from these noisy datasets can help us provide actionable information for assessing risk of a $\mathrm{CO}_2$ storage operation \cite{friedmann2007geological}.
Our aim is to provide information on the state of the geyser by analyzing noisy seismic signals. 
To achieve this goal, we need a framework to differentiate signals that are far in time from an eruption event from signals that are closer to the event under noisy conditions.
An approach to develop such a framework is through machine learning \cite{rouet2017machine,mudunuru2017scalable,holtzman2018machine,rouet2018breaking,wu2018deepdetect,wu2018seismic}.

Machine learning (ML) is a widely used tool for extracting features \cite{christ2018time,kanter2015deep} and classifying seismic signals.
In addition to earthquake detection and prediction \cite{rouet2017machine,rouet2018breaking}, ML is shown to be successful in various other subsurface applications  \cite{mudunuru2017regression,mudunuru2017sequential,vesselinov2018unsupervised,hunter2018reduced,mudunuru2018estimating,marone2018training}.
Within the context of ML, there are many available techniques that are able to predict the evolution of time-series and distinguish seismic signals, within a certain tolerance \cite{de2015grammar,konar2017time}.
However, the main drawback of many ML methods is that predictions are usually affected by noise \cite{li2010trees,mnih2012learning}.
The presence of noise in data is a major concern in classification approaches as the performance of ML classifiers deteriorate \cite{li2010trees,lopez2013insight,saez2016evaluating}.
Interpretability of ML models is hindered by poor classification performance, which can have a negative effect on event discrimination capability.
To overcome this drawback, our novelty is to filter the seismic signal multiple times using high-pass filter and an Autoregressive (AR) model to remove/reduce the effect of different types of noises (temperature effect, human/animal activity, etc).
Then, we classify if a seismic signal is remnant noise or precursor or eruption signal.
The proposed methodology is compared against classical time-series methods such as Dynamic Time Warping (DTW) \cite{xi2006fast,konar2017time} on filtered data and ML methods with partial filtering.
We show that better accuracy can only be achieved through a combination of both filtered data and a robust classification algorithm that is less sensitive to noise.
For ML classification, we use Random Forests (RF), which is a model-free ML method that is less influenced by noise \cite{zhu2004class,saez2016evaluating}. 
A major advantage of model-free ML methods is that they conform to the intrinsic data characteristics with fewer assumptions and without the use of any \textit{a priori} models.
Model-free methods (such as RF) construct non-parametric models using ensembles of multiple base learners without simplifying the underlying problem \cite{breiman2001random,genuer2008random}.

\subsection{Main contributions and outline of the paper}
\label{SubSec:S1_Main_Contributions}
The main contribution of this study is to develop a ML based framework to classify seismic signals under noisy environments and apply it to the Chimay\'{o} geyser eruptions.
An advantage of the proposed method is that even under high anthropogenic noise, we can differentiate the seismic signals that are far away from eruption time to the signals that are closer to the eruption time with an accuracy greater than 90\%.
Moreover, the computational time taken by the proposed ML model to classify a given seismic signal as an eruption or a non-eruption event is in $\mathcal{O}(10^{-4})$ seconds on a laptop.
This makes our ML methodology ideal for usage in eruption event discrimination (such as detecting precursors and differentiating it from background noise) or forecasting geyser eruptions in real-time under noisy conditions.
The paper is organized as follows:~Sec.~\ref{Sec:S2_Chimayo_Method} provides a detailed description of our proposed ML method for distinguishing eruption events from non-eruption events. 
First, we describe the geyser data, the location of sensors, and relevant background on data collection.
Second, we present a method to preprocess the seismic signals to remove temperature effects and background noise.
Third, we construct ML classifiers based on Random Forests for discriminating events on the filtered and partially filtered data.
Sec.~\ref{Sec:S3_Chimayo_Results} discusses the results of the proposed ML method and DTW in classifying signals.
The accuracy of the RF classifiers on filtered and partially filtered data are also provided in this section.
Finally, conclusions are drawn in Sec.~\ref{Sec:S4_Chimayo_Conclusions}.

%% file: Sections/S2_Chimayo_Method.tex

\section{GEYSER DATA AND MACHINE LEARNING METHODOLOGY}
\label{Sec:S2_Chimayo_Method}

\subsection{Site description and data resources}
\label{SubSec:S2_Data_Resources}
Chimay\'{o} geyser located near Chimayo, NM, is one of the few $\mathrm{CO}_2$-driven cold-water geysers in the USA.
It is a man-made geyser and was formed when locals drilled a well for water usage.
There is a shallow aquifer that is beneath the geyser.
The aquifer is within a sub-basin of the Rio Grande Rift, where there is a regional development of $\mathrm{CO}_2$ gas.
Experiments performed on this geyser showed elevated $\mathrm{CO}_2$ conditions at the site \cite{keating2011challenge}.
Moreover, many of the drinking water wells in the neighborhood of this geyser have high levels of dissolved $\mathrm{CO}_2$.
The eruption cycle of the Chimay\'{o} geyser is a simple two-part process.
The system begins with a very long recharge period.
Then, there is a possible bubbling activity and/or minor eruption before major eruption activity \cite{watson2014eruption}. 
The entire minor and major eruption duration averages approximately five minutes.
The eruption dynamics of the geyser is mainly gas-dominated due to a small cross-sectional area of the wellbore \cite{watson2014eruption}.

Fig.~\ref{Fig:Chimayo_Geyser_Sensor_Locations} shows the map of Chimay\'{o} geyser and the location of seismic station RGEYB.
This seismic station is located approximately 3 meters from the well.
The sampling frequency of RGEYB is 200 Hz.
We analyze 18 days of continuous data collected from May-2017 to November-2017 --
14 days of data is used for training and 4 days for testing the ML models.
Eruption images are captured using a motion sensor installed at this station, which are used to label the seismic data. 
The motion sensor records the geysering event time and captures the image when $\mathrm{CO}_2$-rich fluids are ejected out of the wellbore.
The major challenge in analyzing this seismic dataset is that the anthropogenic noise is higher than the seismic signals corresponding to geyser's activity (for example, see Fig.~\ref{Fig:Chimayo_Description}(a)).
The time between eruptions is 17-23 hours while the background noise is all round the clock (generally higher during the day).

\subsection{Preprocessing seismic signals}
\label{SubSec:S2_Preprocessing}
Due to environmental effects (daily temperature variations, seasonal changes, rainfall/thunderstorms, wind etc.) and human/animal activities (cars passing through the driveway, people walking near the geysers, farming, animals running around the geyser etc.), the seismic data contains a lot of noise. 
To achieve higher signal-to-noise ratio (SNR), we filter the seismic signal in two stages.
In the first stage of the filtering process, the Butterworth-Highpass (BH) filter \cite{smith2013digital} is used to remove the seasonal trend.
This filter removes the signals below a certain frequency using corners, thereby allowing higher frequencies to pass through. 
The second stage of filtering uses an AR model to train and predict the anthropogenic noise. 
The reason to use an AR model for denoising is as follows:~In seismic signal processing \cite{robinson2000geophysical}, typically, a band-pass filter is used to extract the signals from a specific frequency band. 
The frequency range is usually selected based on previous research experience and spectrogram analysis \cite{claerbout1964detection}. 
However, in our case, it is not straight-forward to determine the appropriate frequency band of interest directly from the spectrogram of the seismic signal. 
As the precursor in seismic signal has a lower amplitude compared to the eruption signal, a filtering method without prior information would be ideal for our analysis.
We find that a classical time-series forecasting method like an AR model has a good predictability of the ambient noise, which could be used to denoise the seismic signals \cite{lesage2008automatic,kang2012filtering}. 

In the AR modeling, each time-series data point is regressed on its neighborhood values, called the prediction region.
In some specific cases, an AR model can be regarded as a low-pass filter.
An AR model divides the signal into two additive components, a predictable signal and a prediction error signal. 
Let the time-series data obtained from RGEYB seismometer be denoted as $X_t$, where $t = 1, 2, \cdots, n$.
Let $\mathrm{AR}(p)$, be an AR model of order $p$, that estimates $X_t$ using the lagged variable $X_{t-i},~i=1, 2, \cdots, p$, as:~$X_t = c + \sum_{i=1}^p \alpha_i X_{t-i} + \epsilon_t$, where $\alpha_1, \alpha_2, \cdots, \alpha_p$ are the parameters of the model, $c$ is the constant term and $\epsilon_t$ is white noise. 
The parameters of an $\mathrm{AR}(p)$ model are estimated via the ordinary least squares (OLS) procedure.
$\mathrm{AR}(p)$ model coefficients can be thought of as describing the envelope of the spectrum of the seismic signal.
There is no straightforward way to determine the correct model order.
As one increases the value of $p$, the RMS error of an $\mathrm{AR}(p)$ model fit for data generally decreases quickly and then slows down. 
An $\mathrm{AR}(p)$ model order, just after the point at which the RMS error flattens out, is usually an appropriate order.
There are more formal techniques for choosing $p$, the most common of which is the Akaike Information Criterion (AIC) \cite{seabold2010statsmodels}.
In our case, we use AIC to select the value of $p$.
For denoising our seismic signal, we first estimate the $\mathrm{AR}(p)$ model parameters using only 5\% of the entire data (which is mainly noise).
This 5\% of the data does not include the precursor and eruption signals. 
Then, we compute a one-step prediction for the rest of the time-series data. 
The prediction error filter (PEF) of the $\mathrm{AR}(p)$ model is then defined as the original time-series minus the $\mathrm{AR}(p)$ model prediction. 
We apply the PEF filter on the seismic data after BH filtering.
It should be noted that $\mathrm{AR}(p)$ model can only predict stationary noise.
As the geyser dynamics are non-stationary, the corresponding precursor and eruption signals will also be non-stationary.
Hence, removing the stationary part from the original time-series data will not only result in higher SNR but also enhances the precursor and eruption signals in the filtered data.
Algorithm \ref{alg:forecast} and Fig.~\ref{Fig:ML_Filtering_Workflow} summarize the proposed prepossessing step to denoise the seismic signals.
In Sec.~\ref{Sec:S3_Chimayo_Results}, we will show that this multi-stage filtering approach enhances the seismic signal and improves the result of ML classification.

\subsection{Machine learning on seismic signals}
\label{SubSec:ML_Time_Series}
After preprocessing the raw seismic signals, we apply a ML method on the filtered data to distinguish eruption and non-eruption signals. 
Algorithm \ref{alg:forecast} and Fig.~\ref{Fig:ML_Filtering_Workflow} also provide a summary of ML work-flow to classify the state of Chimay\'{o} geyser.
For the classification of seismic signals, the RF method is used.
RF is a popular ensemble ML method that uses many trees to classify unknown signal data by averaging the prediction of each tree.
RF methods, when trained, maintained, and reinforced properly have great potential in solving real-world problems \cite{rouet2017machine,rouet2018breaking}.
They can efficiently process noisy signals through randomization, which is achieved by constructing decision trees in the ensemble using bootstrap samples.
RF selects a random subset of predictor variables at each node to grow the decision tree.
It is most likely that individual decision trees avoid noise contributing input records \cite{genuer2008random}. 
Many variable unbiased decision trees are learned by RF.
In the training dataset, trees that model and overfit noise tend to have a high variance and low bias.
Through majority voting rule, RF reduces the variance, thereby producing a ML model that has a low bias and low variance \cite{breiman2001random}. 
Given the above, it is evident that RF methods could be well suited to deal with noisy seismic signals.

The ML method to classify geyser state starts with feature extraction on a sliding time window to extract a feature vector. 
Then the feature vector is imported into an RF classifier, which acts as a decision-making tool that is capable
of categorizing signals into different classes.
Note that in each time window, there are lots of data points due to the high sampling rate (200 Hz).
Window length assumed is 1 minute, which corresponds to 12000 data points.
Simply using the time-series in the sliding window as the feature vector will lead to the curse of dimensionality.
Hence, we compress the time-series data in a given window into a low dimensional space (which is achieved by extracting features in that window). 
In our case, we calculate over 700 time-series features using \textsf{Tsfresh} Python package \cite{christ2018time}.
The feature vector includes scalar features (e.g., entropy, mean, trend etc.) and vector features (e.g., power spectral density using Welch's method, coefficients of FFT, CWT, ARIMA etc. \cite{christ2018time}).
Feature selection is performed using statistical relevance tests \cite{christ2018time}. 
This type of feature engineering approach reduces the dimension of our time-series data to around 100. 
The top 10 features in the 100 down-selected features include partial autocorrelation lag, FFT coefficients, index mass quantile, time-reversal asymmetry statistic, spectral centroid (mean), variance, skew, and kurtosis of the absolute Fourier transform spectrum, aggregated linear trend, and change quantiles \cite{christ2018time}.
RF model uses these down-selected features instead of each data point in the time window. 

For classification, we label the seismic signal based on the ground truth images.
This results in three label classes.
\texttt{Class-1} correspond to signals that are far away in time from a major eruption event.
That is, any data point in the seismic signal that falls beyond 3 minutes before the eruption event is categorized as \texttt{Class-1}, which can be due to geyser recharge or remnant noise.
\texttt{Class-2} describes the signals that are within 1 minute to 3 minutes before the eruption.
This label corresponds to precursors to a major eruption event.
\texttt{Class-3} represent the signals that fall within the major eruption event, which is 2 minutes.
In this state, the geyser is degassing and releasing $\mathrm{CO_2}$-rich fluids out of the wellbore rapidly.
RF classifier constructed on the training dataset is then applied to distinguish these three classes on the test dataset. 

\begin{algorithm}
  \caption{Overview of proposed ML methodology to classify geyser state from noisy seismic signals}
  \label{alg:forecast}
  \begin{algorithmic}[1]
    \STATE{\textbf{Input}:~Seismic time-series $X_t$ and eruption event times.}
    \STATE{\textbf{Preprocessing}:}
      \begin{itemize}
        \item Apply BH filter with corner/cut-off frequency of $0.1$ on $X_t$ using \textsf{Obspy} Python package. 
          This results in a filtered signal $X^h_t$.
        \item Fit AR($p$) model on 5\% of the filtered data $X^h_t$ using \textsf{statsmodels} Python package.
        \item One-step AR($p$) model prediction $\Tilde{X}^h_t$ on the rest of the 95\% of the data.
        \item Calculate PEF, which is $X^{PEF}_t = X^h_t - \Tilde{X}^h_t$.
      \end{itemize}
    \STATE{\textbf{Multi-class classification}:}
      \begin{itemize}
        \item Seismic time-series labeled into three different classes:
          \begin{itemize}
            \item \texttt{Class-1}:~System is far away from eruption.
            \item \texttt{Class-2}:~System is close to eruption. This state corresponds to possible precursor in seismic signal.
            \item \texttt{Class-3}:~System is in the stage of eruption.
          \end{itemize}
        \item Feature extraction and feature selection on $X^h_t$ and $X^{PEF}_t$ using \textsf{Tsfresh} Python package. 
        \item RF classifier construction on down-selected features on training data. ML analysis performed using \textsf{scikit-learn} Python package. 
      \end{itemize}
    \STATE{\textbf{Output}:~Classification results on test data.} 
      \begin{itemize}
        \item Geyser state is in either \texttt{Class-1} or \texttt{Class-2} or \texttt{Class-3}.
        \item RF classifier accuracy for only BH filtered seismic data, which is $X^h_t$.
        \item RF classifier accuracy for BH filtered + AR($p$) model filtered seismic data, which is $X^{PEF}_t$.
      \end{itemize}
\end{algorithmic}
\end{algorithm}

%% file: Sections/S3_Chimayo_Results.tex

\section{RESULTS}
\label{Sec:S3_Chimayo_Results}
In this section, we present preprocessing and classification results of the proposed ML methodology.
Fig.~\ref{Fig:Chimayo_Seismic_Signals} and Fig.~\ref{Fig:Chimayo_Description}(c) show examples of the recorded signals at RGEYB on days July 7, 2017, and July 11, 2017.
The seismic sensor RGEYB provides 3 components.
Seismic component-1, 2, and 3 correspond to vertical motion of the ground, North-South direction, and East-West direction.
The amplitude of the eruption is relatively small compared to the anthropogenic noise.
Quantitatively, the amplitude of the seismic signal around the major eruption event is approximately less than 1\% of the peak amplitude of the noise signal.
As there are roads, crops, and animal activity around the well, the effect of human noise on the recorded seismic signals is significant.
It should be noted that noise is high from 7 AM to 4 PM.
During this time there is a lot of farming activity and car/truck movement near the geyser.

In this paper, we analyze 18 days of data collected at RGEYB. 
This corresponds to the data collected from May-1-2017 to November-1-2017.
We concentrate on seismic component-1, which is the vertical motion of the ground.
Based on our data analysis, the other two seismic components also showed similar characteristics, and are not shown here for the sake of conciseness.
Also, we note that not much information could be obtained from infrasound signals.
As the seismometer RGEYB is located right next to the geyser (see Fig.~\ref{Fig:Chimayo_Description}(a)), it becomes sensitive to the weather conditions. 
Fig.~\ref{Fig:Temp_Effect}(a) and Fig.~\ref{Fig:Temp_Effect}(c) show an example of the daily trend in the seismic signal.
To construct the daily trend, we calculate the average of seismic amplitude over all the data points sampled in a given second.
Each second of the recorded seismic signal contains 200 raw data points.
The average seismic amplitude is plotted to show the temperature effect on the seismic signal.
The peak in average seismic amplitude is located around 6 AM local time in Chimay\'{o}, NM, which is the coolest hour of the day.
Fig.~\ref{Fig:Temp_Effect}(a) also shows that the time-series in the last few days become less stable.
This is due to the monsoon season in Chimay\'{o}, NM \cite{weatherspark}.
These seasonal trends and temperature effect can be removed via a high-pass filter with appropriate corner frequency.
Fig.~\ref{Fig:Temp_Effect}(b) and Fig.~\ref{Fig:Temp_Effect}(d) show the filtered plots after applying BH filter as discussed in the Sec.~\ref{SubSec:S2_Preprocessing}.
The BH filter is based on \textsf{Obspy} Python package \cite{beyreuther2010obspy} with a corner frequency of 0.1 Hz.
From Fig.~\ref{Fig:Temp_Effect}, we can see that the seasonal trends and temperature effects are removed from the seismic data after high-pass filtering. 

After removing the temperature effect, we train and apply AR($p$) model to reduce the anthropogenic noise from univariate time-series data.
As described in Sec.~\ref{SubSec:S2_Preprocessing}, 5\% of the entire 18 days of data is used to train the AR($p$) model for predicting the anthropogenic noise.
To construct the AR($p$) model, we used \textsf{statsmodels} Python package \cite{seabold2010statsmodels}.
Other inputs or parameters for AR($p$) model are as follows:~OLS solver estimates the AR($p$) model coefficients with convergence tolerance of $10^{-8}$. 
AIC is used for selecting the optimal lag length.
Fig.~\ref{Fig:pef_exp} shows the AR($p$) model prediction of background noise.
We perform one-step prediction of the AR($p$) model on the rest of the data. 
It should be noted that the entire 18 days of seismic data is analyzed.
For illustration purposes, we only show certain representative chunks of time-series that are mean-shifted, stacked, and concatenated together.
From these figures, it is clear that the AR($p$) model predicts the background noise relatively well when compared to the eruption signals with a $R^2$-score of $0.86$ on the remaining 95\% data.
As a result, PEF subtracts the predictions of AR($p$) model from the BH-filtered seismic signal to obtain high SNR data. 

Fig.~\ref{Fig:PEF_RF_Classification} show examples of high SNR signals obtained from the AR($p$) filtering process. 
The results in Fig.~\ref{Fig:pef_exp}(b) and Fig.~\ref{Fig:PEF_RF_Classification} show that PEF largely reduces the variance of the anthropogenic noise while still keeping the precursors and eruption event signals. 
It is also very interesting to see that the same AR($p$) model works well on different eruptions. 
In addition to PEF filtered signals, Fig.~\ref{Fig:PEF_RF_Classification} also provides RF-classifier prediction of geyser state.
From this figure, it is evident that the ground truth and RF-classifier predictions at the end of the classes are close to each other.
Hence, we can infer that RF-method on the PEF filtered signals is able to predict various classes accurately ($\approx 94\%$).
Table~\ref{Tab:Geyser_State_Class_Accuracy_1} provides a summary of classification accuracy of different ML methods to discern geyser state from noisy data.
Three different methods are investigated, which are DTW, RF without PEF, and RF with PEF.
To construct RF-model, we used \textsf{scikit-learn} Python package \cite{scikit-learn}.
14 days (approx. 70\%) of the data is used for training and remaining 4 days (approx. 30\%) is used for testing.
$k$-nearest neighbors with DTW \cite{xi2006fast} is applied for geyser state classification.
The value of $k$ is chosen to be equal to 1.
The number of trees in the random forest ML-model is equal to 100 and bootstrapping is used when building trees.
Gini criterion \cite{breiman2001random} is used to measure the quality of split for the tree nodes.
A minimum number of samples required to be at a leaf node is equal to 1 and all the training samples are given equal weight.
Classification accuracy is based on precision, recall, and $F_1$-score over three classes.
Precision tells us about the positive predictive value.
That is the proportion of positive identifications of geyser states that are actually correct.
Recall provides the proportion of actual positives that are identified correctly.
The $F_1$-score provides the weighted average of the precision and recall.
Classification accuracy reaches the best and worse when $F_1$-score is equal to 1 and 0.

Table.~\ref{Tab:Geyser_State_Class_Accuracy_2} provides the precision, recall and $F_1$-score averaged over three classes.
From these tables, it is evident that DTW with 1-nearest neighbor performs poorly.
The most surprising aspect of classification results is that RF without PEF also performs decently.
Meaning that RF prediction on the partially-filtered seismic data after removing temperature effects has an average $F_1$-score of 0.87.
However, RF classification is enhanced significantly after removing the anthropogenic noise (which is by applying AR($p$) model on the seasonal/temperature filtered seismic data).
From these tables, it is clear that the RF classifier with PEF can distinguish the geyser states better (greater than 90\%) than RF without PEF.
Moreover, the results (see Fig.~\ref{Fig:PEF_RF_Classification}) from the proposed ML approach implies that the time-series near the eruption (\texttt{Class-2}) are substantially different from the time-series data long before the eruption (\texttt{Class-1}). 
The precursors can be bubbling activity, minor eruption or $\mathrm{CO_2}$-rich fluid slowing oozing out of the wellbore or small $\mathrm{CO_2}$-bubbles coalescing to form larger bubbles, which was not previously identified.
Therefore, the inference from this classification study reinforces the fact that precursory signals \cite{rouet2018breaking,rouet2017machine} exist in the seismic data before the major eruption event takes place.

\begin{table}
  \centering
    \caption{Summary of classification accuracy of different ML methods.
    \label{Tab:Geyser_State_Class_Accuracy_1}}
    \resizebox{\textwidth}{!}{\begin{tabular}{|c|c|c|c|c|c|c|c|c|c|} \hline
      \multirow{2}{*}{{\small System/Geyser state}} & 
      \multicolumn{3}{|c|}{{\small DTW}} & 
      \multicolumn{3}{|c|}{{\small RF without PEF}} & 
      \multicolumn{3}{|c|}{{\small RF with PEF}} \\ 
      \cline{2-10}
      & Precision & Recall & $F_1$-score & Precision & Recall & $F_1$-score & Precision & Recall & $F_1$-score \\ \hline
      {\small Class-1} & {\small 0.45} & {\small 1.00} & {\small 0.62} & {\small 0.71} & {\small 1.00} & {\small 0.83} & {\small 0.83} & {\small 1.00} & {\small 0.91}  \\ 
      {\small Class-2} & {\small 0.50} & {\small 0.40} & {\small 0.44} & {\small 1.00} & {\small 0.60} & {\small 0.75} & {\small 1.00} & {\small 0.80} & {\small 0.89}  \\ 
      {\small Class-3} & {\small 1.00} & {\small 0.17} & {\small 0.29} & {\small 1.00} & {\small 1.00} & {\small 1.00} & {\small 1.00} & {\small 1.00} & {\small 1.00}  \\ 
      \hline
    \end{tabular}}
\end{table}

\begin{table}
  \centering
    \caption{Summary of average classification accuracy of different ML methods.
  \label{Tab:Geyser_State_Class_Accuracy_2}}
  \begin{tabular}{|c|c|c|c|} \hline
    Methods & Avg. Precision & Avg. Recall & Avg. $F_1$-score \\ \hline
    DTW & $0.67$ & $0.50$ & $0.44$ \\ 
    RF without PEF & $0.91$ & $0.88$ & $0.87$ \\ 
    RF with PEF & $0.95$ & $0.94$ & $0.94$ \\
    \hline
  \end{tabular}
\end{table}

%% file: Sections/S4_Chimayo_Conclusions.tex

\section{CONCLUDING REMARKS}
\label{Sec:S4_Chimayo_Conclusions}
In this paper, we presented a ML methodology to classify Chimay\'{o} geyser state using noisy seismic signals.
Our method is based on Random Forests combined with Butterworth-Highpass filter and AR($p$) model. 
First, we provided site description, sensor information, and data sources of Chimay\'{o} geyser.
Sensor data analysis and classification of geyser state is focused on the seismic station RGEYB, which is closest to the geyser location.
We also described the types of noises present in the seismic data during the data collection process.
Second, to achieve high SNR, we proposed an approach to preprocess the seismic signals to remove different types of noises.
In the first stage of the filtering process, BH filter is used to remove the seasonal trend and temperature effects.
This filtering method allows higher frequencies to pass through, thereby removing the effects of temperature on the seismic signals.
Then, in the second stage of the filtering process, AR($p$) model is used to remove/reduce the background noise.
AR($p$) model is trained on this noise using 5\% of the seismic data.
Then, it is used for predicting background noise on the remaining 95\% of the data.
The corresponding prediction accuracy of AR($p$) model is greater than 85\%.
Results showed that AR($p$) model largely reduces the variance of the anthropogenic noise while still keeping the precursors and eruption event signals.
Once these noises (seasonal, temperature, animal, and human activities) are removed, we use RF to classify the filtered signals.
Time-series features are extracted and down-selected using statistical relevance tests for RF classification.
14 days of data is used for training and 4 days of data is used for testing the RF-model.
The average classification accuracy of RF-model on the unseen filtered data is greater than 90\%.
Moreover, the event discrimination ML methodology reinforces that precursors in seismic signals exist in the data before the major eruption event takes place.
This demonstrates the capability of the proposed ML approach to distinguish eruption and precursory signals from background noise.
The proposed ML approach is general and is not limited to geyser state classification.
It can readily be applied to datasets to detect anomalies or disruptions or noises in the data, thereby amplifying the signals that may constitute a physical phenomenon not detected in past.
Lastly, the proposed approach streamlines the development of forecasting methods to extract useful and real-time actionable information in noisy environments.

%% file: Sections/Chimayo_Figures.tex

\begin{figure}
  \centering
  \includegraphics[scale=0.75]{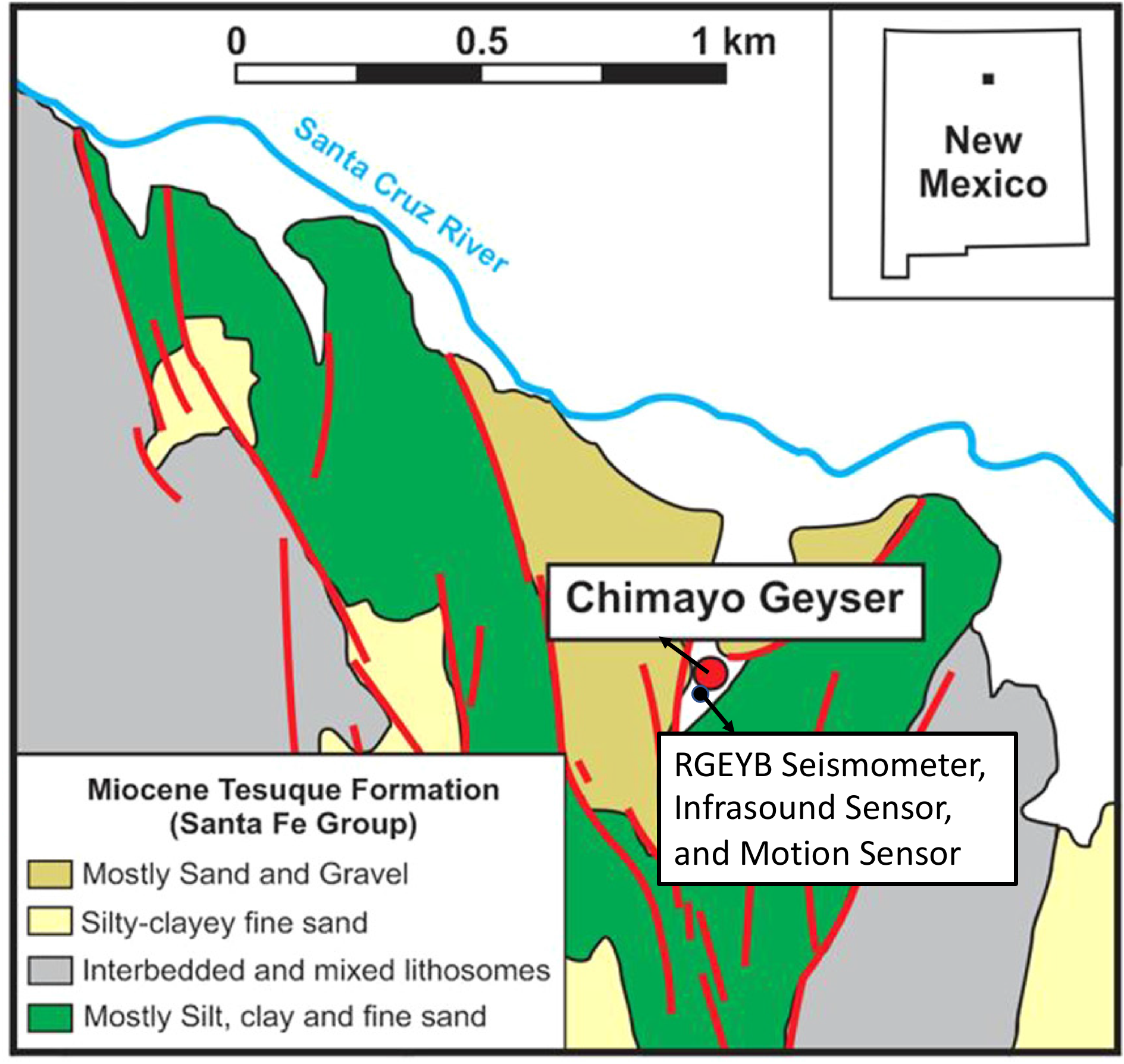}
  \caption{\textsf{\textbf{Chimay\'{o} geyser and sensor locations, New Mexico:}}~This figure shows the map of Chimay\'{o} geyser, New Mexico and sensor locations. 
  Location of Chimay\'{o} geyser is shown as red dot.
  Red lines in this map correspond to faults \cite{keating2010impact,watson2014eruption}.
  The seismic sensor RGEYB is located closest to the geyser and is shown in black dot.
  Infrasound signals are also collected at this seismic station.
  A motion sensor is also installed at RGEYB so that it captures the images of the eruption events.
  These eruption images are then used to create labels for seismic signals collected at RGEYB.
  \label{Fig:Chimayo_Geyser_Sensor_Locations}}
\end{figure}

\begin{figure}
  \centering
  \includegraphics[scale=0.75]{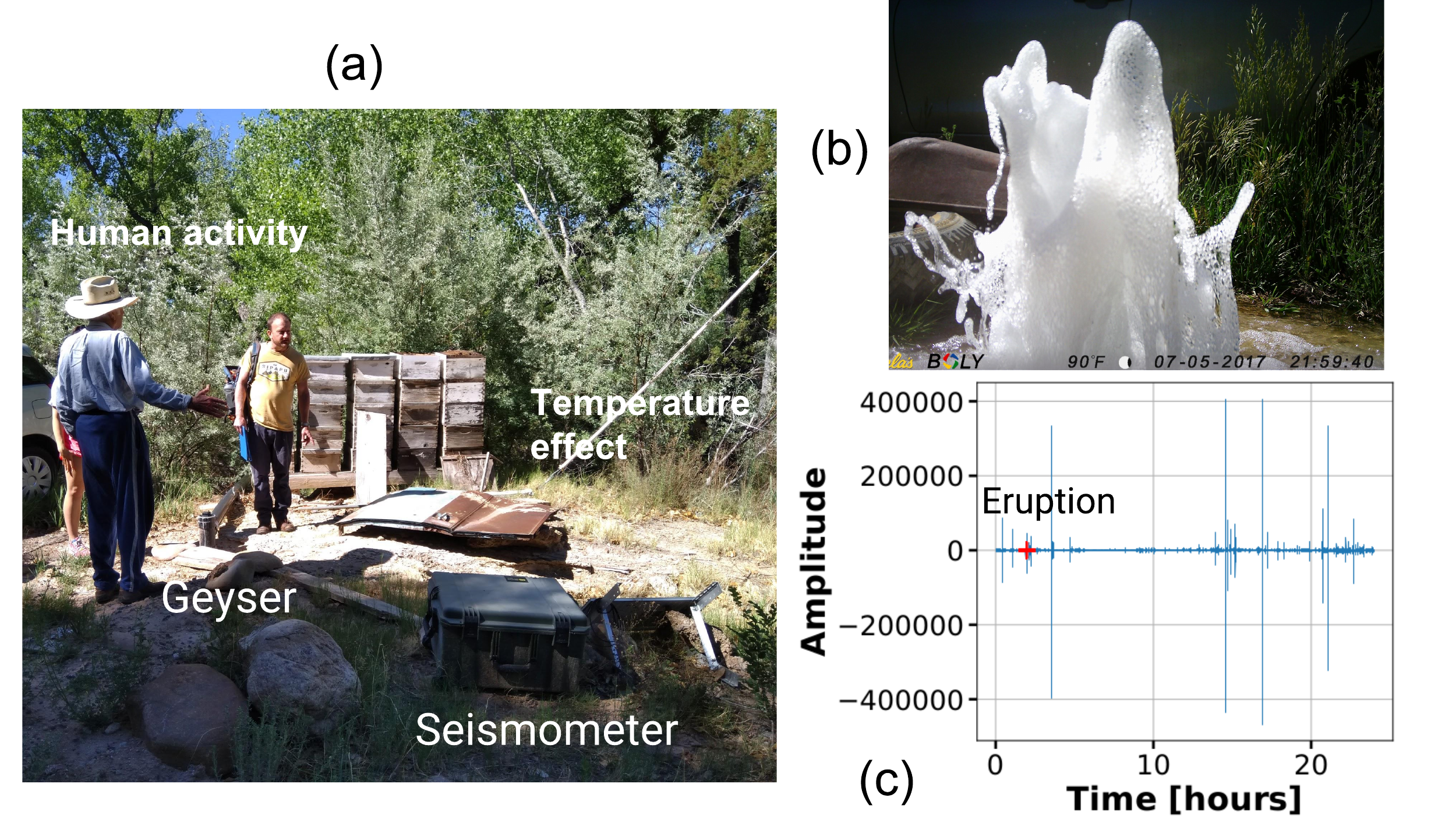}
  \caption{\textsf{\textbf{Chimay\'{o} geyser description:}}~(a) 
  Chimayo geyser after installing the pipe. 
  The seismometer RGEYB and motion sensor are located 3 meters from the geyser.
  Due to its proximity to the local farm and driveway, the effect of ambient noise is significant. 
  (b) The image of a geyser eruption before installing the pipe. 
  (c) A component of seismic signal (amplitude vs. time) recorded on July 11, 2017 at RGEYB seismic station. 
  From this figure, it is clear that the amplitude of the eruption is relatively small compared to the background/anthropogenic noise.
  \label{Fig:Chimayo_Description}}
\end{figure}

\begin{figure}
  \centering
  \includegraphics[scale=0.55]{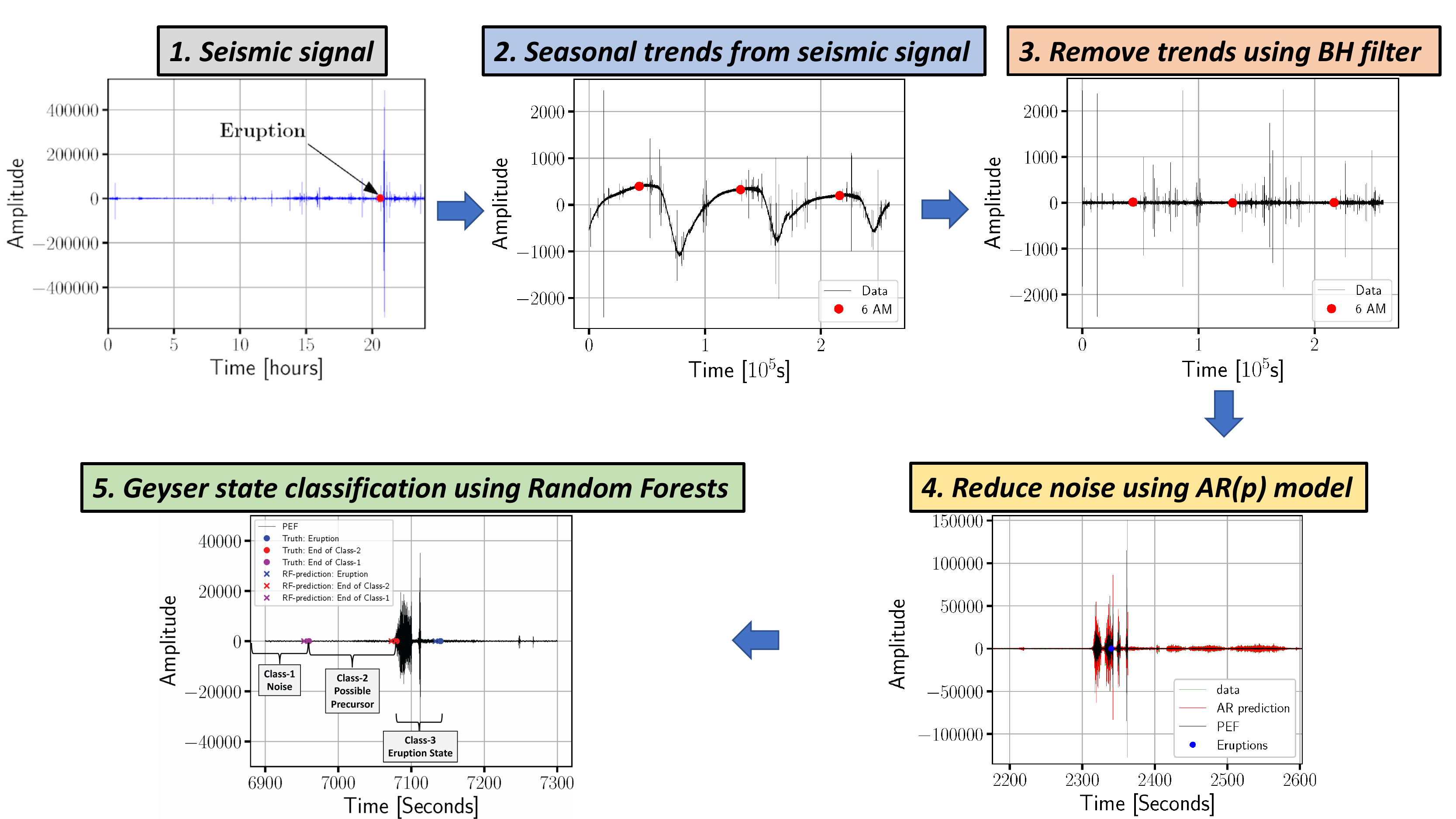}
  \caption{\textsf{\textbf{Noise filtering and ML work-flow:}}~This figure summarizes the proposed noise filtering and ML approach for geyser state classification.
  \label{Fig:ML_Filtering_Workflow}}
\end{figure}

\begin{figure}
  \centering
  \includegraphics[scale=0.925]{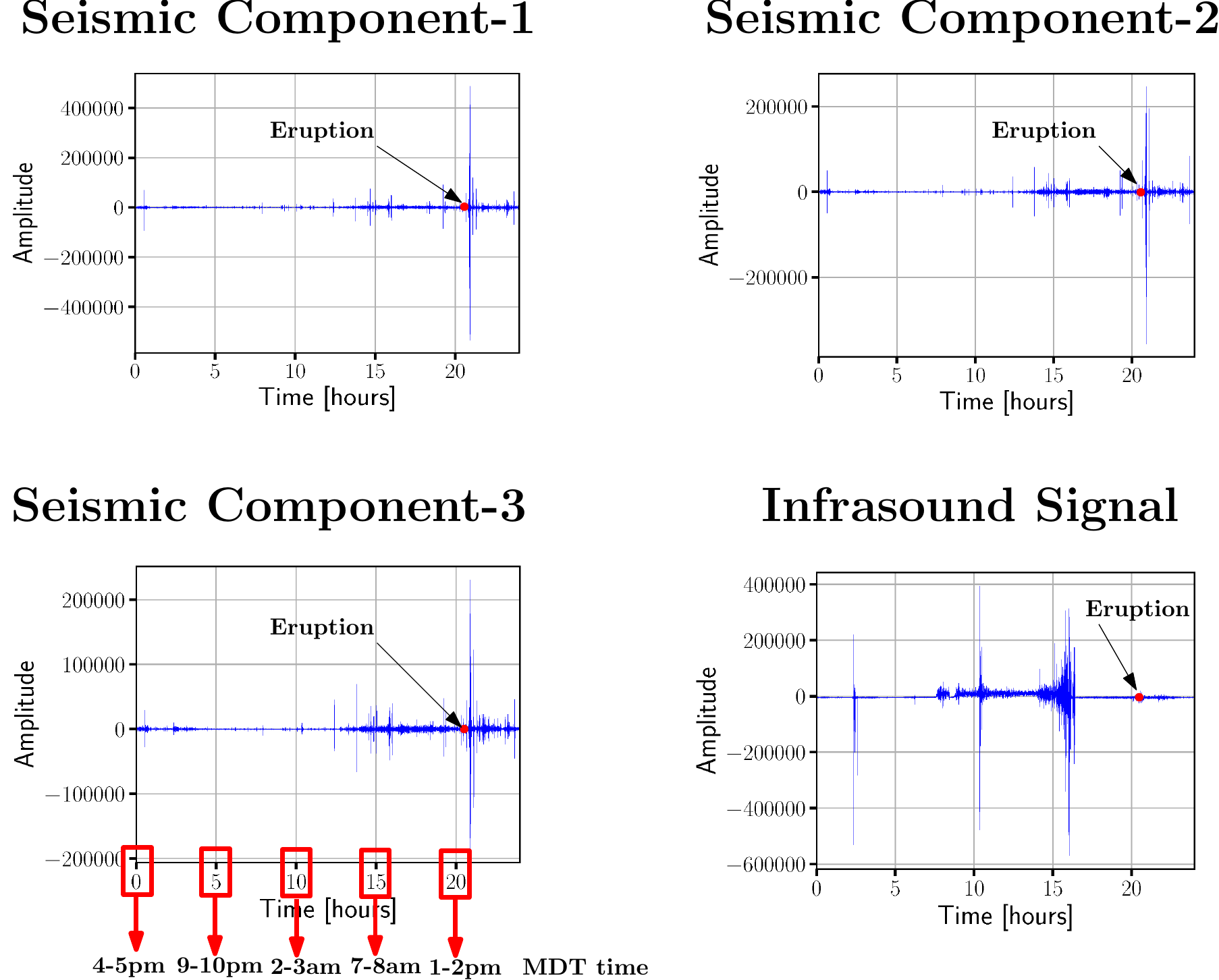}
  \caption{\textsf{\textbf{Examples of seismic and infrasound signals:}}~These figures show the amplitude of seismic and infrasound signals recorded on July 7, 2017 at RGEYB seismic station.
  \label{Fig:Chimayo_Seismic_Signals}}
\end{figure}

\begin{figure}
  \centering
  \subfigure[Seismic signal without BH filtering]
    {\includegraphics[scale=0.50]{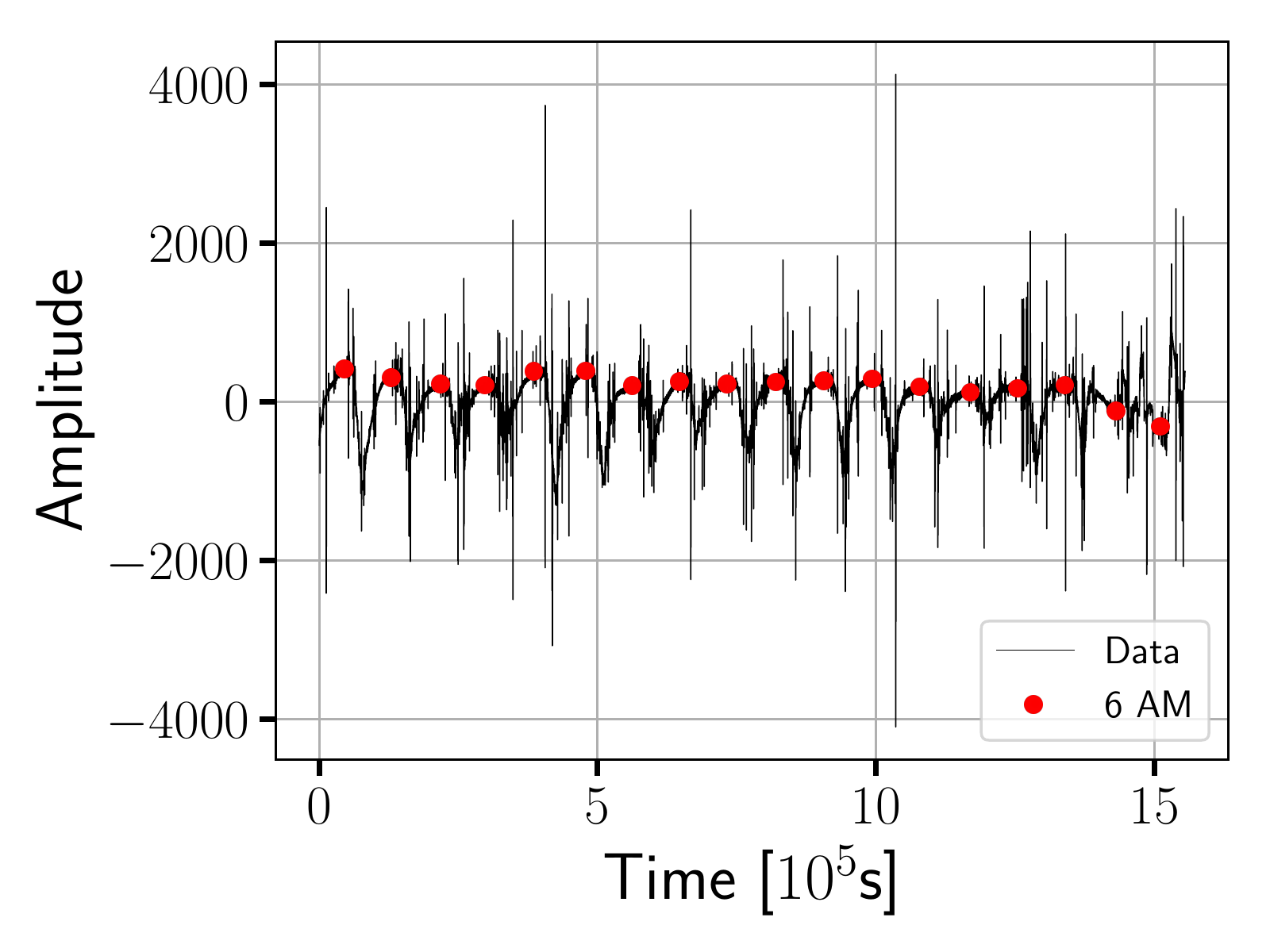}}
  \subfigure[Seismic signal with BH filtering]
    {\includegraphics[scale=0.50]{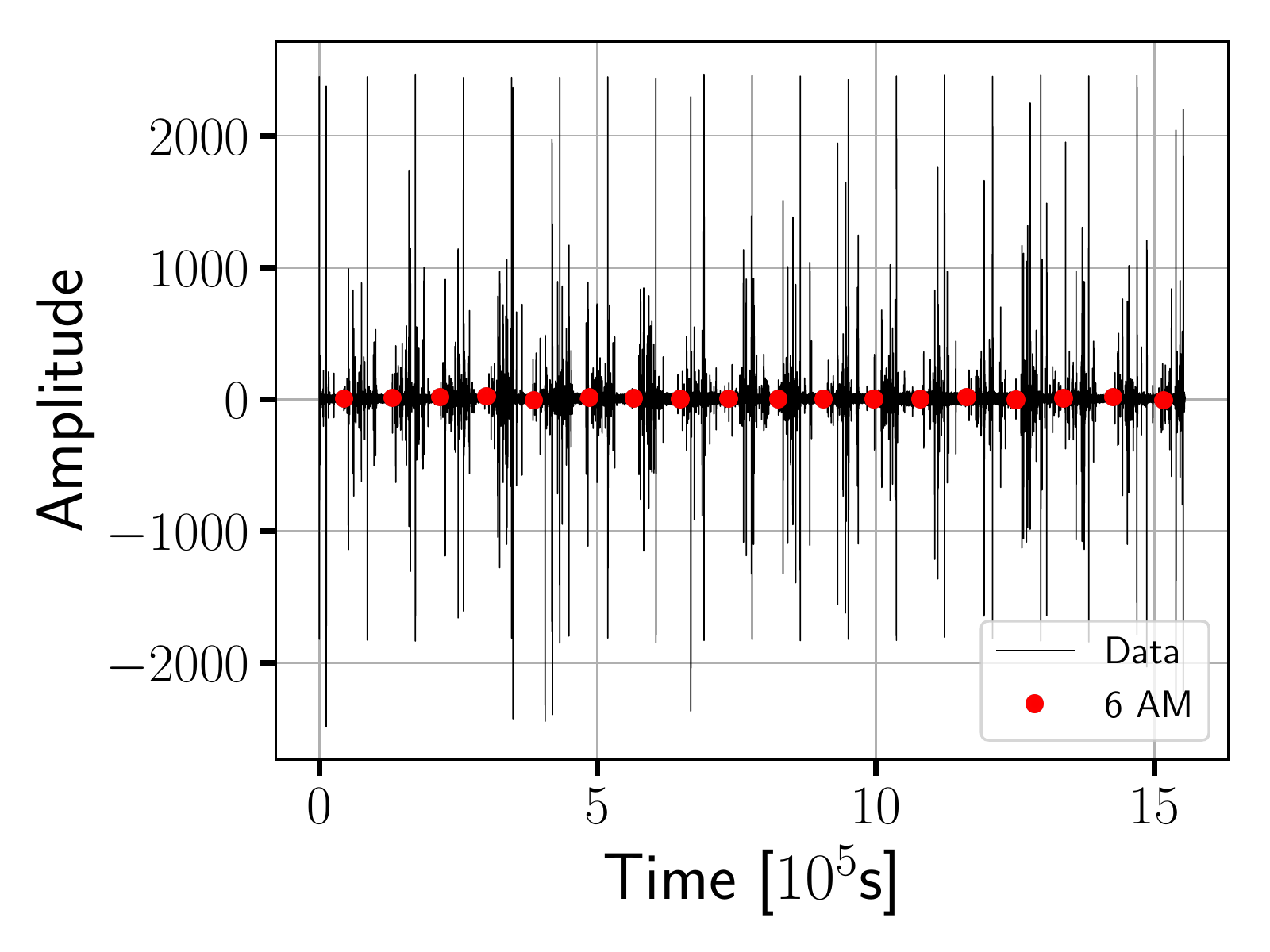}}
  \subfigure[Seismic signal without BH filtering (zoomed in)]
    {\includegraphics[scale=0.50]{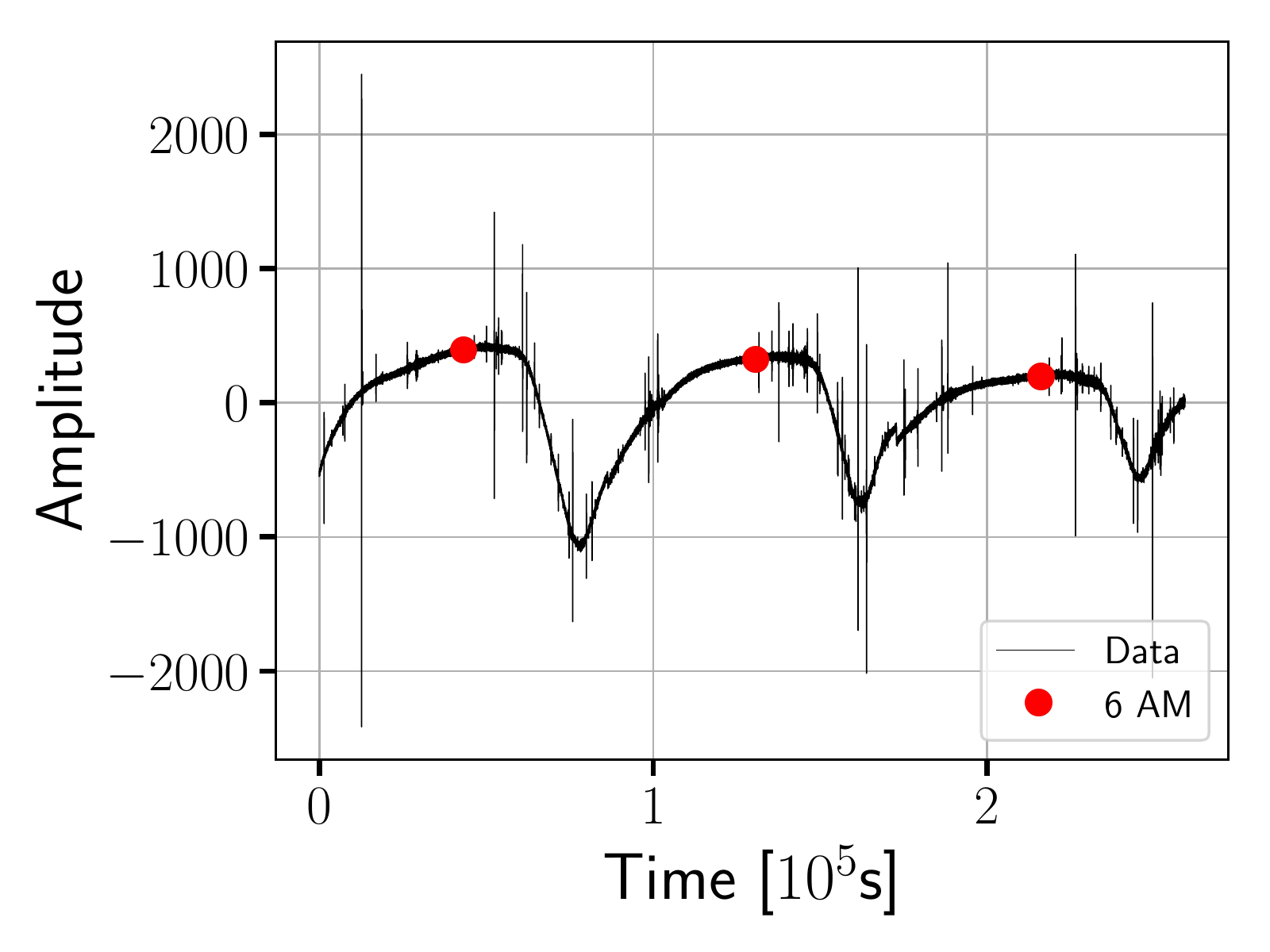}}
  \subfigure[Seismic signal with BH filtering (zoomed in)]
    {\includegraphics[scale=0.50]{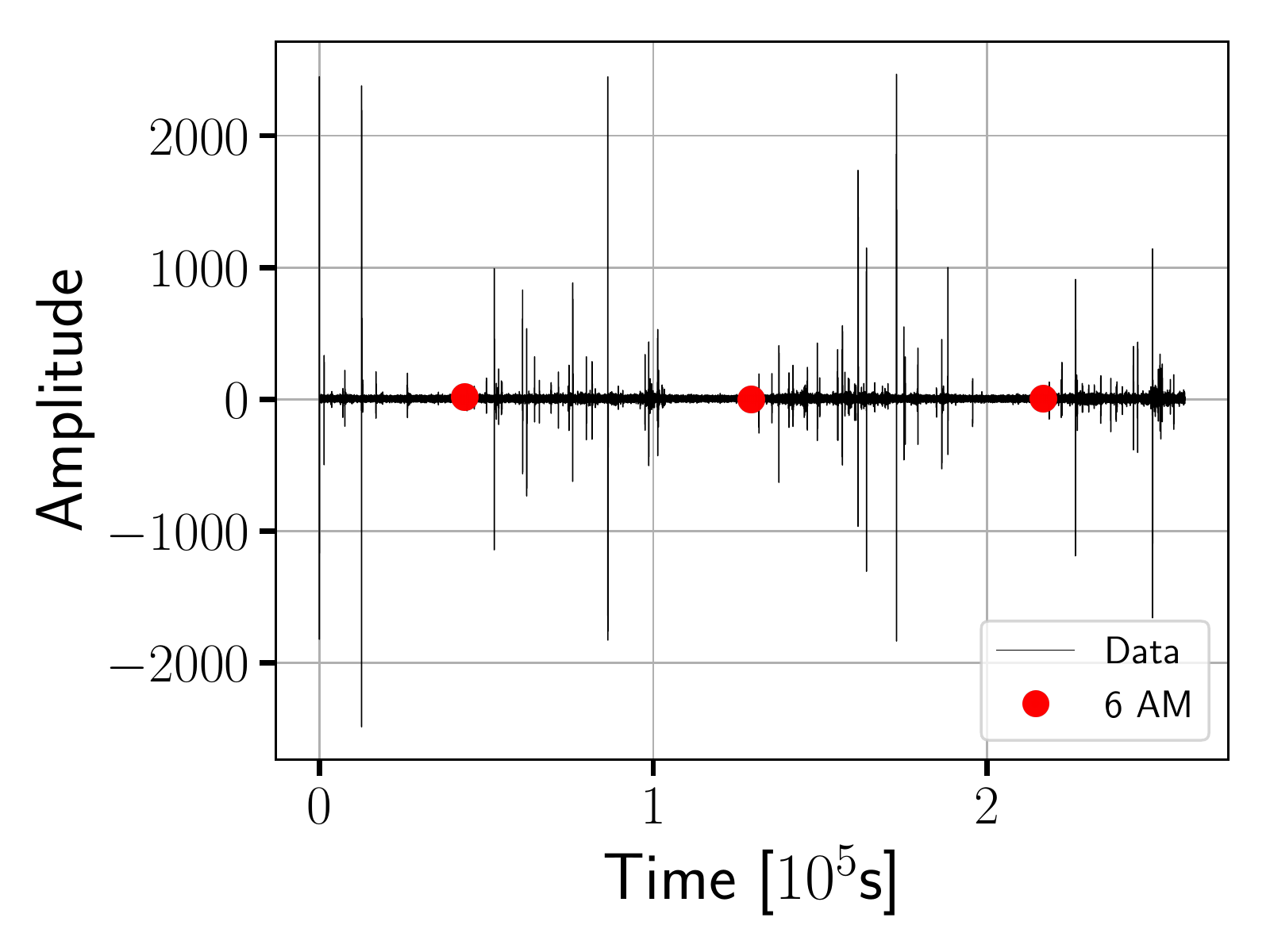}}
  \caption{\textsf{\textbf{Temperature effect on seismic signals and filtering:}}~Figures (a) and (c) show the seasonal trend in the seismic signal.
  Figures (b) and (d) show the filtered plots after applying BH filter on the seismic signal.
  The seasonal trend arising from the temperature is removed from the seismic signals.
  \label{Fig:Temp_Effect}}
\end{figure}

\begin{figure}
  \centering
  \subfigure[AR($p$) prediction on the entire seismic signal]
    {\includegraphics[angle=0, clip, scale=0.65]{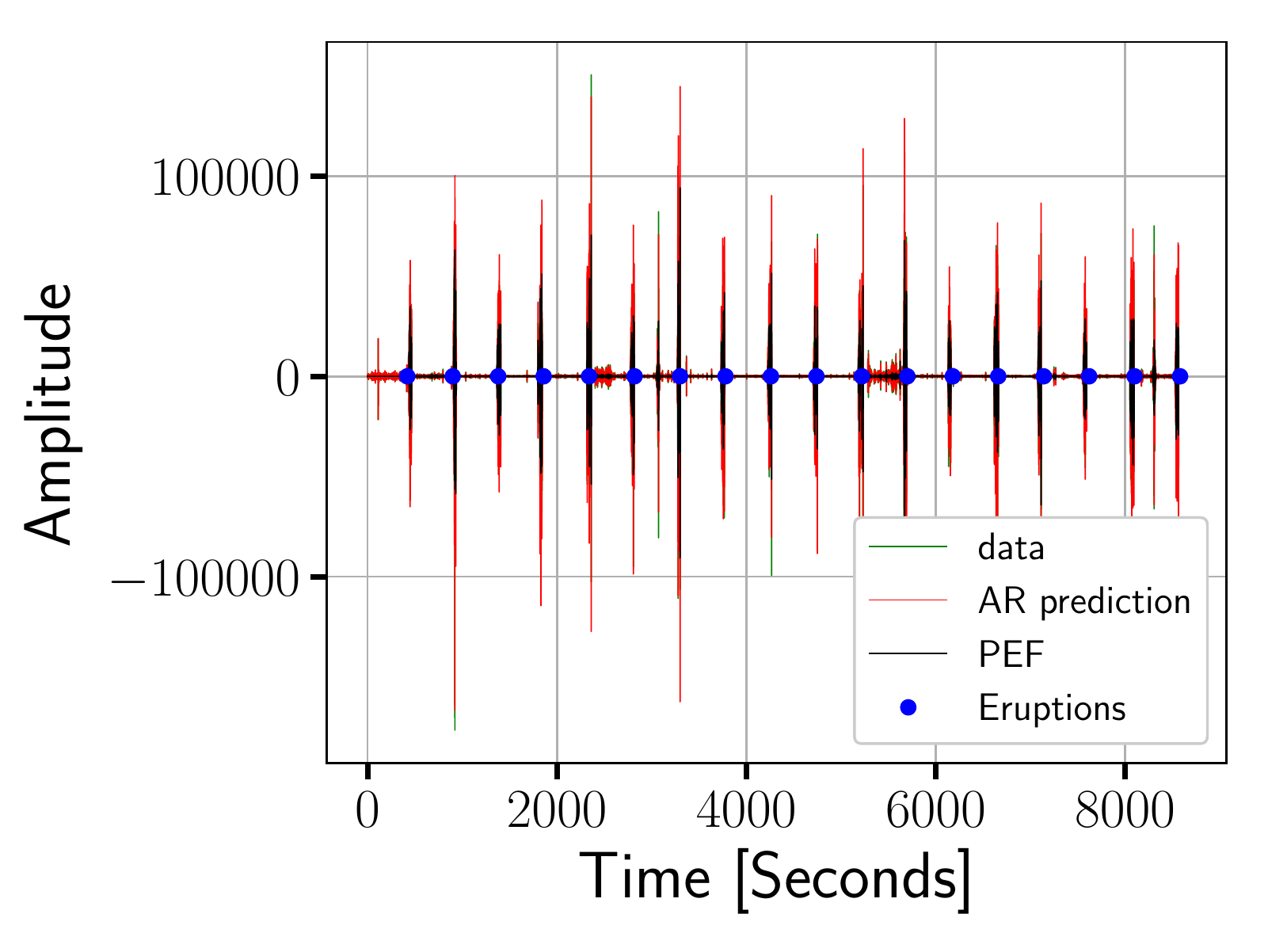}}
  \vspace{-0.15in}
  \subfigure[AR($p$) prediction on the seismic signal corresponding to an eruption event (zoomed-in)]
    {\includegraphics[angle=0, clip, scale=0.65]{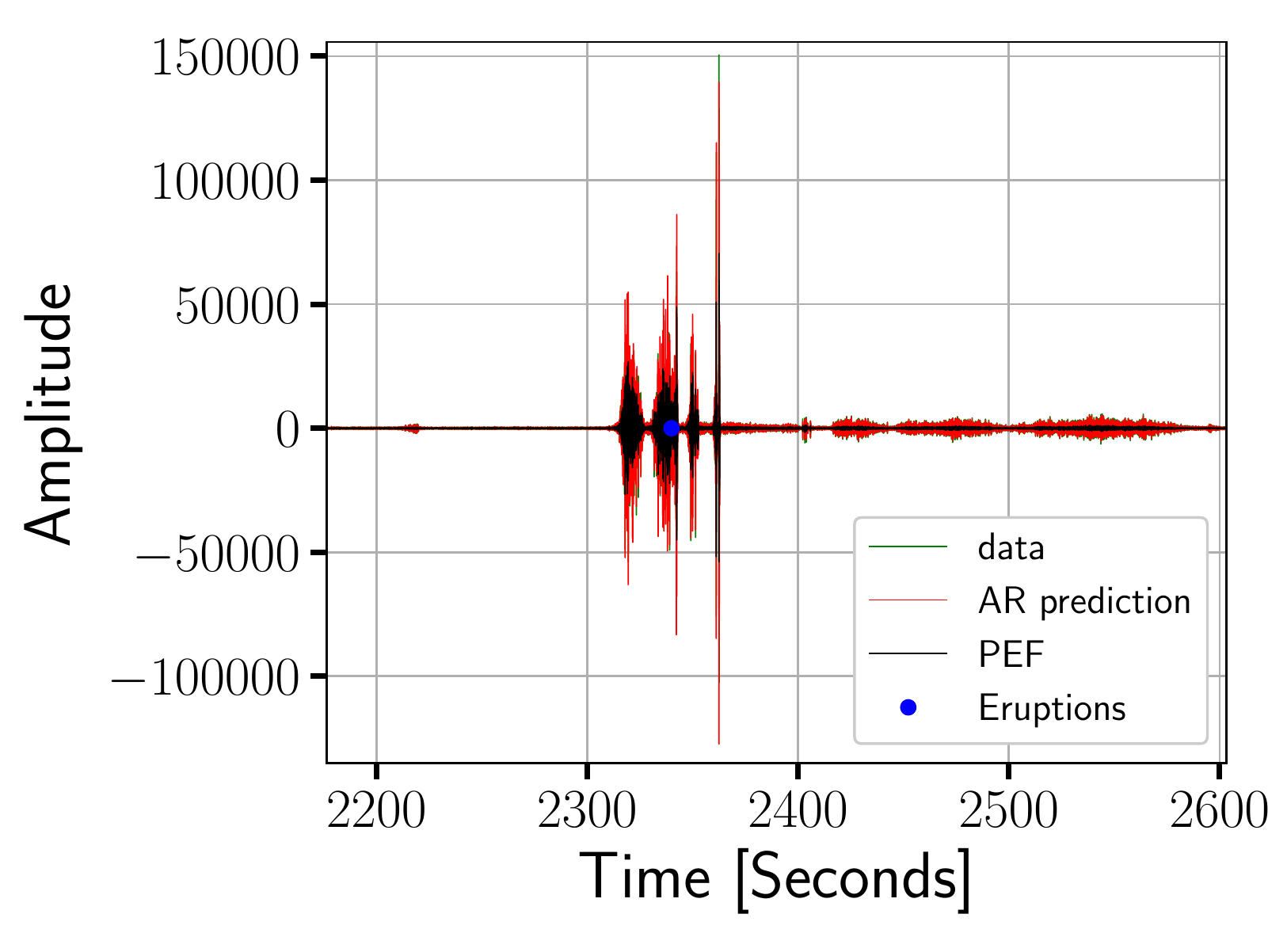}}
  \caption{\textsf{\textbf{Examples of PEF using AR($p$) model:}}~These figures show the AR($p$) prediction of background noise.
  5\% of the entire data set is used for training AR($p$) model and remaining 95\% for noise prediction.
  (a) provides the result of applying PEF filter on the entire seismic signal.
  (b) provides the zoomed in version of (a) near an eruption event.
  The first 5\% data in (a) have no PEF results because the data is used for AR($p$) model training. 
  From the above figures, it is clear that AR($p$) model predicts background noise very well but cannot forecast the eruption signals. 
  PEF (seismic signal after BH filtering - AR($p$) prediction) suppresses the background noise while retaining the precursor signals and eruption event signals.
  \label{Fig:pef_exp}}
\end{figure}

\begin{figure}
  \centering
  \subfigure[Enhanced signals and event classification example]
    {\includegraphics[angle=0, clip, scale=0.55]{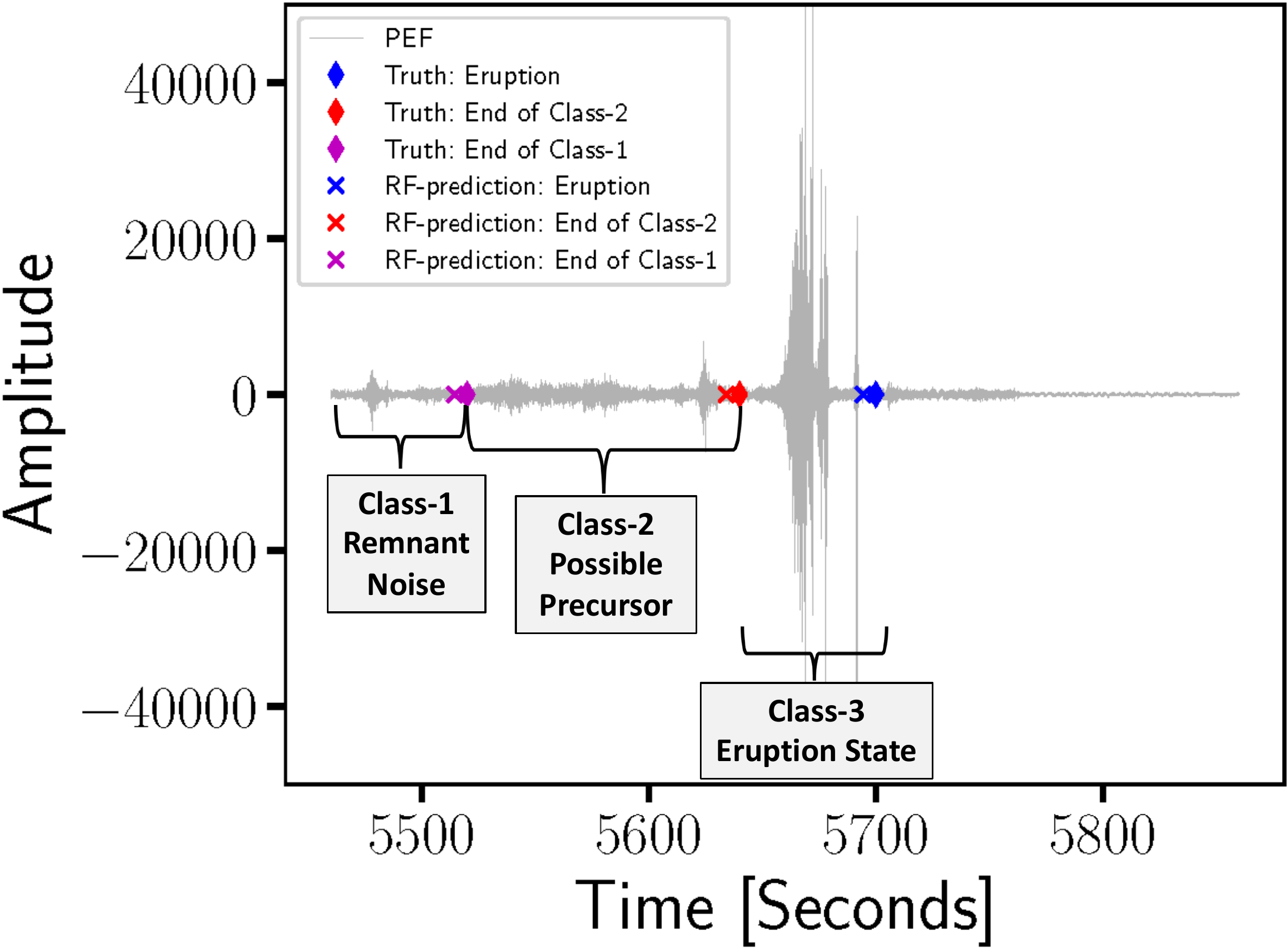}}
  \vspace{-0.15in}
  \subfigure[Enhanced signals and event classification example]
    {\includegraphics[angle=0, clip, scale=0.55]{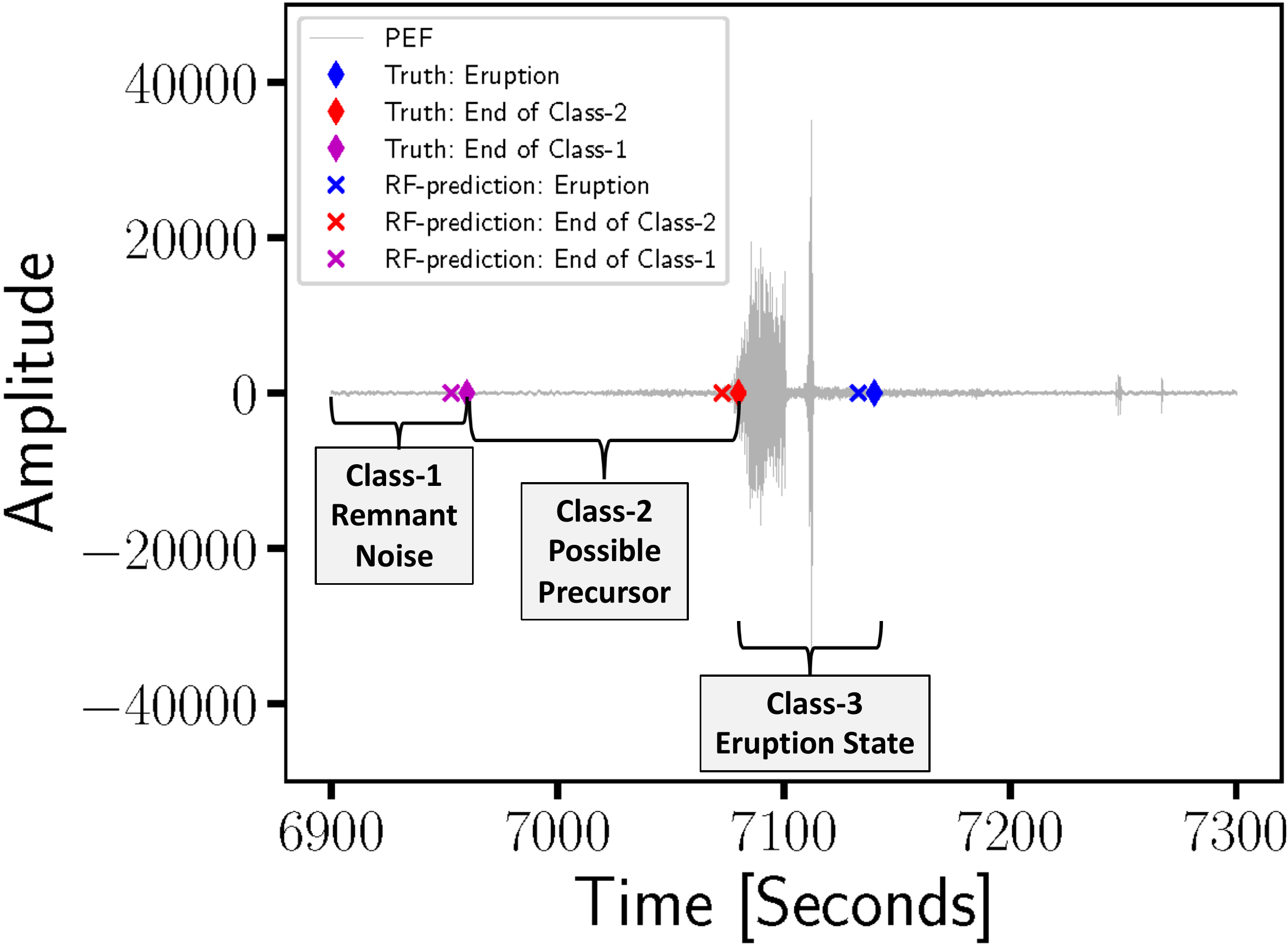}}
  \caption{\textsf{\textbf{High SNR signals after noise filtering and geyser state classification:}}~These figures show examples of noise filtered seismic signals near eruption event.
  It also marks the end of classes corresponding to the ground truth and RF-classifier predictions.
  \label{Fig:PEF_RF_Classification}}
\end{figure}

%% file: Driver_Seismic_Chimayo.bbl
\begin{thebibliography}{10}

\bibitem{carroll2014key}
S.~A. Carroll, E.~H. Keating, K.~Mansoor, Z.~Dai, Y.~Sun, W.~T.-Guitton,
  C.~Brown, and D.~Bacon.
\newblock Key factors for determining groundwater impacts due to leakage from
  geologic carbon sequestration reservoirs.
\newblock {\em International Journal of Greenhouse Gas Control}, 29:153--168,
  2014.

\bibitem{keating2010impact}
E.~H. Keating, J.~Fessenden, N.~Kanjorski, D.~J. Koning, and R.~Pawar.
\newblock The impact of $\mathrm{CO}_2$ on shallow groundwater
  chemistry:~{O}bservations at a natural analog site and implications for
  carbon sequestration.
\newblock {\em Environmental Earth Sciences}, 60:521--536, 2010.

\bibitem{keating2013co2}
E.~H. Keating, J.~A. Hakala, H.~S. Viswanathan, J.~W. Carey, R.~J. Pawar, G.~D.
  Guthrie, and J.~F.-Rahn.
\newblock $\mathrm{CO}_2$ leakage impacts on shallow groundwater:~{F}ield-scale
  reactive-transport simulations informed by observations at a natural analog
  site.
\newblock {\em Applied Geochemistry}, 30:136--147, 2013.

\bibitem{keating2011challenge}
E.~H. Keating, J.~A. Hakala, H.~S. Viswanathan, R.~Capo, B.~Stewart,
  J.~Gardiner, G.~D. Guthrie, J.~W. Carey, and J.~F.-Rahn.
\newblock The challenge of predicting groundwater quality impacts in a
  $\mathrm{CO}_2$ leakage scenario:~{R}esults from field, laboratory, and
  modeling studies at a natural analog site in {N}ew {M}exico, {USA}.
\newblock {\em Energy Procedia}, 4:3239--3245, 2011.

\bibitem{manga2006seismic}
M.~Manga and E.~Brodsky.
\newblock Seismic triggering of eruptions in the far field:~{V}olcanoes and
  geysers.
\newblock {\em Annual Review of Earth and Planetary Sciences}, 34:263--291,
  2006.

\bibitem{hurwitz2017fascinating}
S.~Hurwitz and M.~Manga.
\newblock The fascinating and complex dynamics of geyser eruptions.
\newblock {\em Annual Review of Earth and Planetary Sciences}, 45, 2017.

\bibitem{glennon2005operation}
J.~A. Glennon and R.~M. Pfaff.
\newblock The operation and geography of carbon dioxide-driven, cold-water
  geysers.
\newblock {\em GOSA Transactions}, 9:184--192, 2005.

\bibitem{watson2014eruption}
Z.~T. Watson, W.~S. Han, E.~H. Keating, N.-H. Jung, and M.~Lu.
\newblock Eruption dynamics of $\mathrm{CO}_2$-driven cold-water
  geysers:~{C}rystal, {T}enmile geysers in {U}tah and {C}himay{\'o} geyser in
  {N}ew {M}exico.
\newblock {\em Earth and Planetary Science Letters}, 408:272--284, 2014.

\bibitem{friedmann2007geological}
S.~J. Friedmann.
\newblock Geological carbon dioxide sequestration.
\newblock {\em Elements}, 3:179--184, 2007.

\bibitem{rouet2017machine}
B.~R.-Leduc, C.~Hulbert, N.~Lubbers, K.~Barros, C.~J. Humphreys, and P.~A.
  Johnson.
\newblock Machine learning predicts laboratory earthquakes.
\newblock {\em Geophysical Research Letters}, 44:9276--9282, 2017.

\bibitem{mudunuru2017scalable}
M.~K. Mudunuru, V.~K. Chillara, S.~Karra, and D.~N. Sinha.
\newblock Scalable time-series feature engineering framework to understand
  multiphase flow using acoustic signals.
\newblock In {\em Proceedings of Meetings on Acoustics 6ICU}, volume~32, page
  055003, 2017.

\bibitem{holtzman2018machine}
B.~K. Holtzman, A.~Pat{\'e}, J.~Paisley, F.~Waldhauser, and D.~Repetto.
\newblock Machine learning reveals cyclic changes in seismic source spectra in
  {G}eysers geothermal field.
\newblock {\em Science Advances}, 4:eaao2929, 2018.

\bibitem{rouet2018breaking}
B.~R.-Leduc, C.~Hulbert, and P.~A. Johnson.
\newblock Breaking {C}ascadia's silence:~{M}achine learning reveals the
  constant chatter of the megathrust.
\newblock {\em arXiv preprint arXiv:1805.06689}, 2018.

\bibitem{wu2018deepdetect}
Y.~Wu, Y.~Lin, Z.~Zhou, D.~C. Bolton, J.~Liu, and P.~A. Johnson.
\newblock {D}eep{D}etect:~{A} cascaded region-based densely connected network
  for seismic event detection.
\newblock {\em IEEE Transactions on Geoscience and Remote Sensing}, (99):1--14,
  2018.

\bibitem{wu2018seismic}
Y.~Wu, Y.~Lin, Z.~Zhou, and A.~Delorey.
\newblock {S}eismic-{N}et:.~{A} deep densely connected neural network to detect
  seismic events.
\newblock {\em arXiv preprint arXiv:1802.02241}, 2018.

\bibitem{christ2018time}
M.~Christ, N.~Braun, J.~Neuffer, and A.~W. K.-Liehr.
\newblock Time {S}eries {F}eatu{R}e {E}xtraction on basis of {S}calable
  {H}ypothesis tests (tsfresh--{A} {P}ython package).
\newblock {\em Neurocomputing}, 2018.

\bibitem{kanter2015deep}
J.~M. Kanter and K.~Veeramachaneni.
\newblock Deep {F}eature {S}ynthesis:~{T}owards automating data science
  endeavors.
\newblock In {\em IEEE International Conference on Data Science and Advanced
  Analytics (DSAA)}, pages 1--10, 2015.

\bibitem{mudunuru2017regression}
M.~K. Mudunuru, S.~Karra, D.~R. Harp, G.~D. Guthrie, and H.~S. Viswanathan.
\newblock Regression-based reduced-order models to predict transient thermal
  output for enhanced geothermal systems.
\newblock {\em Geothermics}, 70:192--205, 2017.

\bibitem{mudunuru2017sequential}
M.~K. Mudunuru, S.~Karra, N.~Makedonska, and T.~Chen.
\newblock Sequential geophysical and flow inversion to characterize fracture
  networks in subsurface systems.
\newblock {\em Statistical Analysis and Data Mining:~The ASA Data Science
  Journal}, 10:326--342, 2017.

\bibitem{vesselinov2018unsupervised}
V.~V. Vesselinov, M.~K. Mudunuru, S.~Karra, D.~O. Malley, and B.~S. Alexandrov.
\newblock Unsupervised machine learning based on non-negative tensor
  factorization for analyzing reactive-mixing.
\newblock {\em arXiv preprint arXiv:1805.06454}, 2018.

\bibitem{hunter2018reduced}
A.~Hunter, B.~A. Moore, M.~K. Mudunuru, V.~T. Chau, R.~L. Miller, R.~B. Tchoua,
  C.~Nyshadham, S.~Karra, D.~O. Malley, E.~Rougier, H.~S. Viswanathan, and
  G.~Srinivasan.
\newblock Reduced-order modeling through machine learning approaches for
  brittle fracture applications.
\newblock {\em arXiv preprint arXiv:1806.01949}, 2018.

\bibitem{mudunuru2018estimating}
M.~K. Mudunuru, N.~Panda, S.~Karra, G.~Srinivasan, V.~T. Chau, E.~Rougier,
  A.~Hunter, and H.~S. Viswanathan.
\newblock Estimating failure in brittle materials using graph theory.
\newblock {\em arXiv preprint arXiv:1807.11537}, 2018.

\bibitem{marone2018training}
C.~Marone.
\newblock Training machines in {E}arthly ways.
\newblock {\em Nature Geoscience}, 11:301, 2018.

\bibitem{de2015grammar}
A.~M. De~Silva and P.~H.~W. Leong.
\newblock {\em Grammar-based Feature Generation for Time-Series Prediction}.
\newblock SpringerBriefs in Computational Intelligence. Springer, Singapore,
  2015.

\bibitem{konar2017time}
A.~Konar and D.~Bhattacharya.
\newblock {\em Time-Series Prediction and Applications:~A Machine Intelligence
  Approach}, volume 127 of {\em Intelligent Systems Reference Library}.
\newblock Springer, Cham, Switzerland, 2017.

\bibitem{li2010trees}
H.~B. Li, W.~Wang, H.~W. Ding, and J.~Dong.
\newblock Trees weighting random forest method for classifying high-dimensional
  noisy data.
\newblock In {\em 2010 IEEE 7th International Conference on e-Business
  Engineering (ICEBE)}, pages 160--163, 2010.

\bibitem{mnih2012learning}
V.~Mnih and G.~E. Hinton.
\newblock Learning to label aerial images from noisy data.
\newblock In {\em Proceedings of the 29th International Conference on Machine
  Learning (ICML-12)}, pages 567--574, 2012.

\bibitem{lopez2013insight}
V.~L{\'o}pez, A.~Fern{\'a}ndez, S.~Garc{\'\i}a, V.~Palade, and F.~Herrera.
\newblock An insight into classification with imbalanced data:~{E}mpirical
  results and current trends on using data intrinsic characteristics.
\newblock {\em Information Sciences}, 250:113--141, 2013.

\bibitem{saez2016evaluating}
J.~A. S{\'a}ez, J.~Luengo, and F.~Herrera.
\newblock Evaluating the classifier behavior with noisy data considering
  performance and robustness:~{T}he equalized loss of accuracy measure.
\newblock {\em Neurocomputing}, 176:26--35, 2016.

\bibitem{xi2006fast}
X.~Xi, E.~Keogh, C.~Shelton, L.~Wei, and C.~A. Ratanamahatana.
\newblock Fast time series classification using numerosity reduction.
\newblock In {\em Proceedings of the 23rd International Conference on Machine
  Learning}, pages 1033--1040, 2006.

\bibitem{zhu2004class}
X.~Zhu and X.~Wu.
\newblock Class noise vs. attribute noise:~{A} quantitative study.
\newblock {\em Artificial Intelligence Review}, 22:177--210, 2004.

\bibitem{breiman2001random}
L.~Breiman.
\newblock Random forests.
\newblock {\em Machine learning}, 45:5--32, 2001.

\bibitem{genuer2008random}
R.~Genuer, J.-M. Poggi, and C.~Tuleau.
\newblock Random {F}orests:~{S}ome methodological insights.
\newblock {\em arXiv preprint arXiv:0811.3619}, 2008.

\bibitem{smith2013digital}
S.~W. Smith.
\newblock {\em Digital Signal Processing:~{A} Practical Guide for Engineers and
  Scientists}.
\newblock Elsevier, Burlington, Massachusetts, USA, 2013.

\bibitem{robinson2000geophysical}
E.~A. Robinson and S.~Treitel.
\newblock {\em Geophysical Signal Analysis}.
\newblock Society of Exploration Geophysicists, Tulsa, Oklahoma, USA, 2000.

\bibitem{claerbout1964detection}
J.~F. Claerbout.
\newblock Detection of p-waves from weak sources at great distances.
\newblock {\em Geophysics}, 29:197--211, 1964.

\bibitem{lesage2008automatic}
P.~Lesage.
\newblock Automatic estimation of optimal autoregressive filters for the
  analysis of volcanic seismic activity.
\newblock {\em Natural Hazards and Earth System Science}, 8:369--376, 2008.

\bibitem{kang2012filtering}
E.~L. Kang and J.~Harlim.
\newblock Filtering nonlinear spatio-temporal chaos with autoregressive linear
  stochastic models.
\newblock {\em Physica D:~Nonlinear Phenomena}, 241:1099--1113, 2012.

\bibitem{seabold2010statsmodels}
S.~Seabold and J.~Perktold.
\newblock Statsmodels:~{E}conometric and statistical modeling with python.
\newblock In {\em Proceedings of the 9th Python in Science Conference},
  volume~57, page~61. SciPy society Austin, 2010.

\bibitem{weatherspark}
Average {W}eather in {C}himay\'{o} {N}ew {M}exico, {U}nited {S}tates.
\newblock {\em
  https://weatherspark.com/y/3509/Average-Weather-in-Chimayo-New-Mexico-United-States-Year-Round}.

\bibitem{beyreuther2010obspy}
M.~Beyreuther, R.~Barsch, L.~Krischer, T.~Megies, Y.~Behr, and J.~Wassermann.
\newblock Obs{P}y:~{A} {P}ython toolbox for seismology.
\newblock {\em Seismological Research Letters}, 81:530--533, 2010.

\bibitem{scikit-learn}
F.~Pedregosa, G.~Varoquaux, A.~Gramfort, V.~Michel, B.~Thirion, O.~Grisel,
  M.~Blondel, P.~Prettenhofer, R.~Weiss, V.~Dubourg, J.~Vanderplas, A.~Passos,
  D.~Cournapeau, M.~Brucher, M.~Perrot, and E.~Duchesnay.
\newblock Scikit-learn:~{M}achine {L}earning in {P}ython.
\newblock {\em Journal of Machine Learning Research}, 12:2825--2830, 2011.

\end{thebibliography}
